\documentclass[10pt,journal]{IEEEtran}
\usepackage{graphicx}
\usepackage{amsmath, amsfonts}
\usepackage{xcolor}
\usepackage{amssymb}
\usepackage{algorithm}
\usepackage{algpseudocode}
\usepackage[caption=false,font=footnotesize]{subfig}
\algrenewcommand\alglinenumber[1]{\scriptsize #1}
\algrenewcommand\algorithmicindent{1.5em}
\usepackage{cite}
\usepackage{balance}
\usepackage{setspace}
\usepackage{url}
\usepackage{tikz}
\usetikzlibrary{arrows.meta,positioning,fit,calc}

\ifCLASSINFOpdf

\else

\fi

\begin{document}

\title{Communication-Aware and Safety-Aware UAV Control via Predictive Latent Models}

\author{Hamid Shiri and Mehdi Bennis 
\thanks{Hamid Shiri and Mehdi Bennis are with the Centre for Wireless Communications, University of Oulu, FI-90014 Oulu, Finland(e-mail: \{hamid.shiri,mehdi.bennis\}@oulu.fi).}}

\maketitle

\begin{abstract}

This article presents a communication-aware and risk-aware predictive latent control (CRPL) framework for unmanned aerial vehicle (UAV) systems operating under partial observability and uncertain environment dynamics. CRPL integrates a joint-embedding predictive architecture (JEPA) with probabilistic communication and safety constraints to jointly optimize UAV motion and transmission power. The learned latent model generates recursive multi-step rollouts, enabling the controller to anticipate future motion, channel degradation, and collision risk. These predictions are incorporated into a unified safety-aware optimization framework for proactive, energy-aware trajectory and communication adaptation. Simulation results show that CRPL closely approaches the performance of an oracle analytical predictive controller and outperforms reactive constrained and unconstrained baselines under limited bandwidth and dynamic uncertainty. In the bandwidth-limited regime, CRPL reduces terminal error, i.e., the final UAV-to-goal distance, by up to a factor of approximately $3$ and outage duration by up to approximately $18$, while also lowering communication energy and collision risk. These improvements are achieved with only a moderate motion-energy overhead, demonstrating a favorable trade-off among mobility effort, communication reliability, and operational safety.

\end{abstract}


\IEEEpeerreviewmaketitle

\section{Introduction}
\label{sec:introduction}

Unmanned aerial vehicles (UAVs) are expected to play an important role in future intelligent wireless and autonomous systems because of their mobility, flexibility, and rapid deployment capability \cite{mozaffari2019tutorial,zeng2019accessing}. In applications such as wireless coverage, aerial surveillance, intelligent transportation, emergency communications, environmental monitoring, and edge intelligence, UAVs must simultaneously accomplish mission objectives, maintain reliable connectivity, use communication resources efficiently, and operate safely in uncertain environments \cite{mozaffari2019tutorial,mozaffari2016efficient,MassiveUAVMFG, zhang2019uav,liu2020uav,FedMeanFieldUAV}.

\begin{figure}
  \includegraphics[width=\linewidth, trim=2cm 0cm 2cm 0cm, clip]{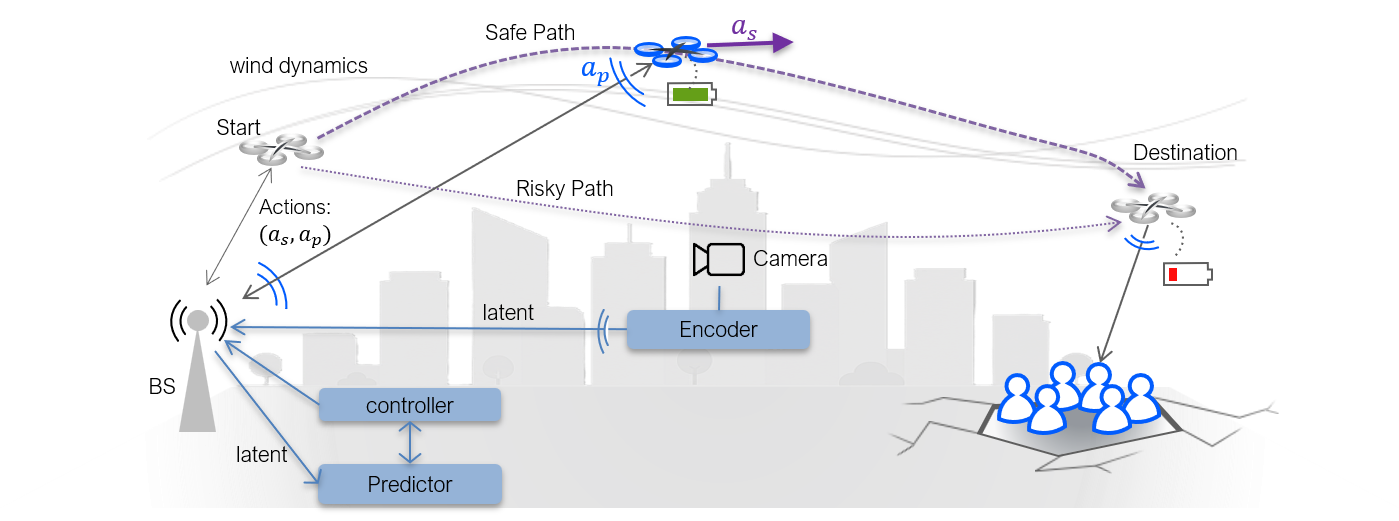}
\caption{Communication-aware UAV control scenario in which a BS controls a UAV from its initial position toward a goal region while accounting for wireless connectivity and obstacle-induced risk.}
  \label{fig:scenario}
  \vspace{-1.5em}
\end{figure}

This problem is illustrated in \figurename{~\ref{fig:scenario}}, where a base station (BS) generates navigation and communication decisions for a UAV using latent representations extracted from high-dimensional sensor observations. Because communication quality and collision risk depend on future interactions between the UAV and its environment, effective operation requires anticipating both channel and safety conditions. This motivates predictive decision-making mechanisms that jointly account for communication reliability, environment evolution, and safety constraints.

A central challenge in autonomous UAV operation is the strong coupling among mobility, communication reliability, environment geometry, and safety. Communication links in UAV scenarios are highly sensitive to blockage and the resulting large-scale fading effects  \cite{khawaja2019survey,AlHourani2014}. Existing communication-aware UAV optimization methods have investigated UAV placement, throughput maximization, outage reduction, coverage enhancement, secure communications, and energy-efficient trajectory design \cite{alzenad20183,wu2018joint,zeng2017energy,zhang2019trajectory}. More recently, learning-based communication-aware control frameworks have been proposed for online UAV path planning and large-scale UAV coordination, including neural-network-based opportunistic control, attention-based communication-aware planning, federated-learning-assisted control, and mean-field game formulations for massive UAV systems  \cite{RemoteUAVOpportunisticControl,MassiveUAVMFG,AttentionBasedCommControl,FedMeanFieldUAV}. Despite these advances, most existing approaches adapt communication and navigation decisions using current observations or instantaneous channel conditions. As a result, their ability to anticipate future communication degradation or collision risks caused by obstacles and environmental uncertainty remains limited.

Meanwhile, advances in model-based reinforcement learning and latent world models have shown that predictive dynamics can be learned directly from high-dimensional observations \cite{hafner2020dream,schrittwieser2020mastering}. Joint-embedding predictive architectures (JEPAs) and their extensions, including I-JEPA and V-JEPA2, provide self-supervised mechanisms for learning predictive latent representations without requiring computationally expensive pixel-level reconstruction \cite{lecun2022path,VJEPA2}. Related studies have explored latent representation learning in wireless systems, including latent wireless dynamics from channel-state information, multi-modal latent dynamics, and predictive latent planning \cite{LatentWirelessDynamics,LatentMultimodalPlanning}. These developments suggest that compact latent representations can capture the geometric and dynamic information required for predictive control. However, their use in joint communication, motion, and safety optimization for UAV systems remains largely unexplored.

Robust and safety-aware control is also challenging under uncertain dynamics, partial observability, and communication degradation. Chance-constrained planning and stochastic model predictive control provide principled mechanisms for enforcing probabilistic safety requirements \cite{blackmore2011chance,mesbah2016stochastic}. Reach-avoid formulations and safe reinforcement learning have further extended these ideas to learning-based decision making under explicit safety constraints \cite{ReachAvoidLearning,Achiam2017,Wabersich2023}. However, these methods typically emphasize navigation safety without jointly predicting communication reliability, while latent-model approaches rarely incorporate both probabilistic communication and collision constraints. A unified framework that combines predictive latent environment modeling, communication-aware planning, and safety-aware UAV control is therefore still lacking.

Motivated by these challenges, this article proposes CRPL, a communication-aware and risk-aware predictive latent control framework for joint UAV trajectory and transmission-power optimization. CRPL combines a JEPA-based latent model with probabilistic communication and safety constraints to generate multi-step predictions of future motion, channel conditions, and collision risk from high-dimensional observations. These predictions enable proactive communication and trajectory adaptation rather than purely reactive control. Simulation results show that CRPL approaches the performance of an oracle analytical predictive controller and improves
terminal accuracy, communication efficiency, reliability, and safety relative to reactive baselines. Although the selected collision-safe and communication-aware trajectories may require moderately greater motion effort, motion effort is explicitly regularized within the objective, yielding an energy-aware trade-off among mobility, reliability, and safety.

The remainder of this article is organized as follows. Section~II formulates the JEPA-based communication-control problem. Section~III presents the safety-aware policy-optimization framework. Section~IV reports the simulation results, and Section~V concludes the article.

\section{System Model}
\label{sec:system_model}

Consider the communication-aware UAV control scenario illustrated in \figurename{~\ref{fig:scenario}}. The environment contains both controllable and uncontrollable entities. The controllable entity corresponds to the UAV, whereas the uncontrollable entities represent obstacles or other objects whose motions cannot be controlled by the BS. The environment state at time step $k$ is represented by
$
s_k=[r_k^T,v_k^T]^T,
$
where $r_k$ and $v_k$ denote the positions and velocities of all entities, respectively, and $(\cdot)^T$ denotes the transpose operation. The analytical environment dynamics are assumed unavailable and are represented abstractly as
\begin{align}
s_{k+1}\sim p_s(\cdot|s_k,a_{s,k}),
\label{Eq:dynamics}
\end{align}
where $a_{s,k}$ denotes the UAV control action and $p_s(\cdot)$ is an unknown stochastic transition model.

\subsection{JEPA-Based Latent Model}
\label{sec:jepa}

\begin{figure}
\centering
\includegraphics[width=0.95\linewidth]{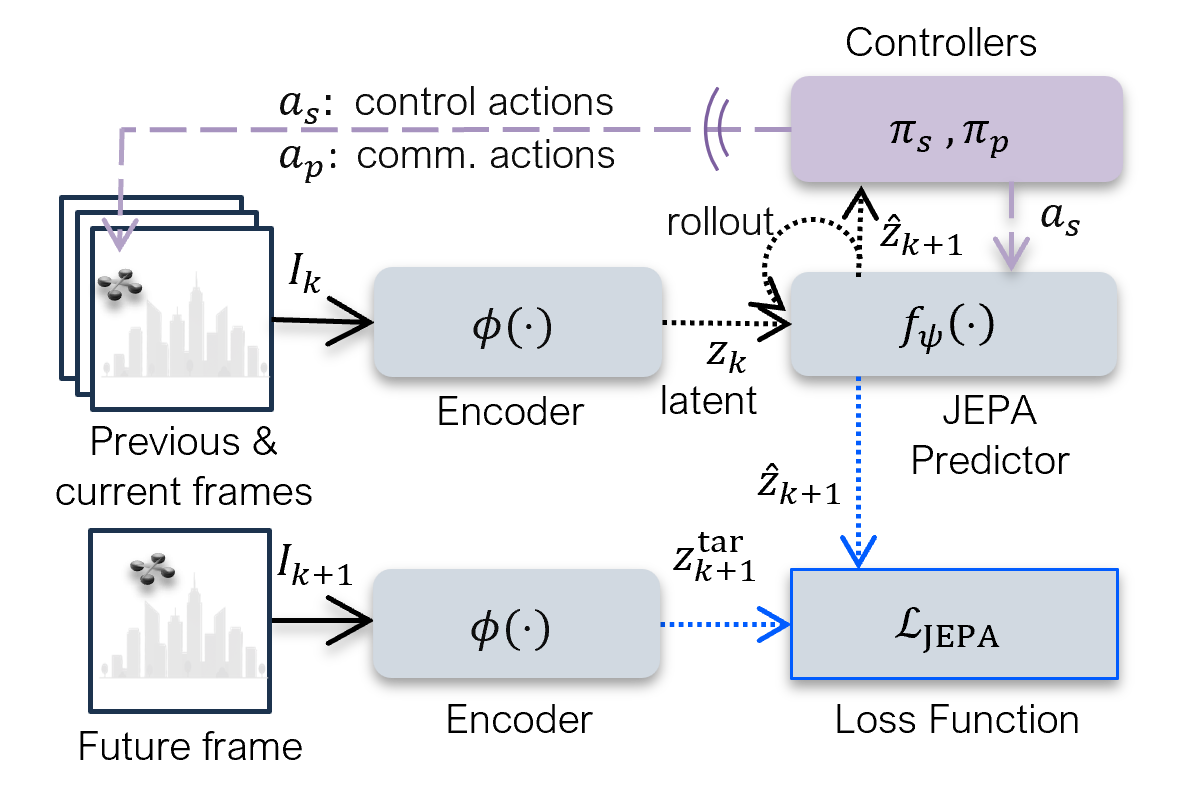}
\vspace{-0.5em}
\caption{JEPA-based latent model for predictive communication-aware UAV control. Future latent states are generated using the current latent representation and control actions.}
\label{fig:scenario2}
  \vspace{-1.5em}

\end{figure}

To model the unknown dynamics in \eqref{Eq:dynamics}, we employ a joint embedding predictive architecture (JEPA) that learns predictive latent representations from high-dimensional observations. At each time step, observations are encoded into a compact latent state, and a latent predictor utilizes the current latent representation together with the actions to estimate future latent states. The predictor is trained by matching its predictions to latent representations generated by a target encoder, thereby learning environment dynamics without explicit system identification. The overall flow of observations, latent representations, communication actions, control actions, and training structure is illustrated in \figurename{~\ref{fig:scenario2}}.

\subsubsection{Latent Representation}

At each time step $k$, the environment is observed through an image $I_k$, as illustrated in \figurename{~\ref{fig:scenario2}}, and encoded into a latent representation
\begin{align}
\label{Eq:representation}
z_k=\phi(I_k),
\end{align}
where $\phi(\cdot)$ denotes the JEPA encoder. The observations are generated by the underlying physical state $s_k$ and provide only a partial description of the environment.

The latent representation captures information associated with both the UAV and its surrounding environment. Following structured and object-centric representation learning approaches \cite{kipf2019contrastive,locatello2020object}, we conceptually partition the latent representation as
\begin{align}
z_k=
\begin{bmatrix}
z_{c,k}^{T},\;
z_{u,k}^{T}
\end{bmatrix}^{T},
\end{align}
where $z_{c,k}$ and $z_{u,k}$ denote the latent components associated with controllable and uncontrollable entities, respectively. This decomposition facilitates the formulation of goal-reaching objectives and safety constraints, while naturally extending to environments containing multiple controllable and uncontrollable entities.

\subsubsection{Latent Dynamics Predictor}

While the latent representation in \eqref{Eq:representation} captures the current environment state, communication- and safe-aware control requires anticipating future environment evolution. To incorporate temporal information, the predictor operates on a history of latent representations. Defining the augmented latent state
\begin{align} \label{Eq:z_k}
\bar z_k=
\begin{bmatrix}
z_k^T,
z_{k-1}^T,
\cdots,
z_{k-M+1}^T
\end{bmatrix}^{T},
\end{align}
where $M$ denotes the history length, the latent predictor generates
\begin{align}
\hat z_{k+1}
=
f_{\psi}(\bar z_k,a_{s,k}),
\label{eq:jepa_predictor}
\end{align}
where $f_\psi(\cdot)$ denotes the JEPA latent predictor and $a_{s,k}$ is the control action. The predicted latent representation $\hat z_{k+1}$ is subsequently appended to the latent history and used to recursively generate future latent trajectories over a planning horizon. For notational simplicity, predicted latent states generated during rollout and policy optimization are denoted by $z_k$ whenever no ambiguity arises.

\subsubsection{JEPA Training}
\label{sec:Jepa_train}

The JEPA encoder and latent predictor are trained offline using data collected from interactions with the environment under exploratory control actions. The objective is to learn latent representations that preserve the predictive structure of the environment while avoiding computationally expensive pixel-level reconstruction. Let
\begin{align}
z_{k+1}^{\mathrm{tar}}
=
\phi_{\mathrm{tar}}(I_{k+1})
\end{align}
denote the target latent representation generated by a target encoder $\phi_{\mathrm{tar}}(\cdot)$ from the future observation $I_{k+1}$. The target encoder shares the same architecture as the online encoder and is updated using an exponential moving average
\begin{align}
\phi_{\mathrm{tar}}
\leftarrow
\tau\,\phi_{\mathrm{tar}}
+
(1-\tau)\,\phi,
\end{align}
where $\tau\in(0,1)$ is the momentum parameter. This slowly evolving target network improves training stability and mitigates representation collapse.

Given the latent history $\bar z_k$ and control action $a_{s,k}$, the predictor generates the latent estimate $\hat z_{k+1}$ according to \eqref{eq:jepa_predictor}. The predictor is trained by minimizing the discrepancy between the predicted latent representation and the target embedding, yielding the JEPA objective
\begin{align}
\mathcal L_{\mathrm{JEPA}}
=
\mathbb E
\!\left[
\left\|
\hat z_{k+1}
-
z_{k+1}^{\mathrm{tar}}
\right\|_2^2
\right].
\end{align}

Unlike reconstruction-based latent models, the JEPA objective directly encourages predictive consistency in latent space. 
Therefore, the learned latent representation captures environmental dynamics that are most relevant to forecasting future observations and downstream decision-making. After training, the encoder $\phi(\cdot)$ and latent predictor $f_\psi(\cdot)$ are fixed and jointly constitute the learned world model used for latent rollout, constraint evaluation, and policy optimization.

\subsection{Problem Formulation}

We formulate a communication-aware and safety-aware UAV control problem over a finite planning horizon $H$. At each time step $k$, the controller samples a motion action $a_{s,k}\sim\pi_s(\cdot|\bar z_k)\in\mathcal A_s$ and a communication-power action $a_{p,k}\sim\pi_p(\cdot|\bar z_k)\in\mathcal A_p$, where $\mathcal A_s$ and $\mathcal A_p$ denote the corresponding feasible action spaces, $\pi_s$ and $\pi_p$ denote the control and communication policies, respectively, and $\bar z_k$ denotes the augmented latent state defined in \eqref{Eq:z_k}.

\subsubsection{Control Objective}
The objective is to drive the controllable latent state $z_{c,k}$ toward a desired goal representation $z_g$ while minimizing both control effort and communication power consumption. Accordingly, the stage cost is defined as
\begin{align}
\ell_k
=
\| z_{c,k}-z_g\|_2^2
+
\lambda_s\| a_{s,k}\|_2^2
+
\lambda_p\| a_{p,k}\| _2^2 ,
\end{align}
where $\lambda_s>0$ and $\lambda_p>0$ determine the relative importance of control and communication energy expenditures.
The communication-aware and safety-aware requirements are enforced through the following chance constraints.

\subsubsection{Collision Avoidance Constraint}

To ensure safe operation, the probability of collision with uncontrollable entities should remain sufficiently low throughout the planning horizon. Let
$
\mathbf z_c^H
=
\{z_{c,1},z_{c,2},\ldots,z_{c,H}\}
$
and
$
\mathbf z_u^H
=
\{z_{u,1},z_{u,2},\ldots,z_{u,H}\}
$
denote the controllable and uncontrollable latent trajectories, respectively. The safety requirement is imposed through the chance constraint
\begin{align}
\Pr_{\pi_s,\pi_p}
\Big(
\mathbf z_c^H
\in
\mathcal{S}_{\mathrm{safe}}
(
\mathbf z_u^H
)
\Big)
\ge
1-\delta_1,
\label{eq:safe_traj_constraint}
\end{align}
where $\delta_1\in[0,1)$ specifies the allowable collision risk.

The trajectory-level safety set is defined as

\vspace{-10pt}
\small
\begin{align}
\mathcal{S}_{\mathrm{safe}}
(
\mathbf z_u^H
)
=
\Big\{
\mathbf z_c^H
~\Big|~
\Omega_{c,k}
\cap
\Omega_{u,k}
=
\emptyset,
~
\forall k=1,\ldots,H
\Big\},
\end{align}
\normalsize

\noindent
where $\Omega_{c,k}$ and $\Omega_{u,k}$ denote the observed occupancy regions of the controllable and uncontrollable entities in the observation $I_k$ at time step $k$, respectively.
The condition
$
\Omega_{c,k} \cap \Omega_{u,k} =\emptyset
$
ensures that the controllable and uncontrollable entities occupy disjoint image regions throughout the planning horizon, thereby indicating collision-free operation.

\subsubsection{Communication Reliability Constraint}
Reliable wireless connectivity should be maintained throughout the planning horizon. Communication quality is quantified by the received signal-to-noise ratio (SNR), which depends on the UAV's location, surrounding obstacles, communication power, and the BS's position. Accordingly, the received SNR at time step $k$ is modeled as

\vspace{-10pt}
\small
\begin{align}
\gamma_k =\mathrm{SNR}\left(\Omega_{c,k},a_{p,k},\Omega_{u,k},s_{\mathrm{BS}}\right) \nonumber \\
=\frac{\left|h\left(\Omega_{c,k},\Omega_{u,k},s_{\mathrm{BS}}\right)\right|^2a_{p,k}}{P_n},
\label{Eq:snr_def}
\end{align}
\normalsize

\noindent
where $h(\cdot)$ denotes the wireless channel coefficient, $s_{\mathrm{BS}}$ represents the BS state, and $P_n$ is the noise power. The channel coefficient depends on the relative geometry of the UAV, obstacles, and BS, thereby capturing blockage effects and geometry-dependent channel variations.

To ensure reliable communication, the probability that the received SNR remains above a minimum threshold $\gamma_{\min}$ throughout the planning horizon should exceed a prescribed reliability level. To express communication reliability over the horizon, let $\mathbf a_p^H= \{a_{p,1},a_{p,2},\ldots,a_{p,H} \}$ denote the communication-power trajectory. The corresponding received-SNR trajectory is $\boldsymbol{\gamma}^{H} \left(\mathbf z_c^H,\mathbf z_u^H,\mathbf a_p^H,s_{\mathrm{BS}}\right) = \{\gamma_1,\gamma_2,\ldots,\gamma_H \}$. For notational convenience, we subsequently write $\boldsymbol{\gamma}^{H}=\boldsymbol{\gamma}^{H} \left(\mathbf z_c^H,\mathbf z_u^H,\mathbf a_p^H,s_{\mathrm{BS}}\right)$. The trajectory-level communication reliability set is then defined as
\begin{align}
\mathcal{S}_{\mathrm{comm}} =\left\{\boldsymbol{\gamma}^{H}~\middle|~\gamma_k\ge \gamma_{\min},~\forall k=1,\ldots,H\right\}.
\end{align}
The communication reliability requirement is enforced through the chance constraint
\begin{align}
\Pr_{\pi_s,\pi_p}\Big(\boldsymbol{\gamma}^{H}\in \mathcal{S}_{\mathrm{comm}}\Big)\ge1-\delta_2,\label{eq:snr_constraint}
\end{align}
where $\delta_2\in[0,1)$ specifies the maximum allowable outage probability.

\subsubsection{Optimization Problem}
Combining the  cost, collision avoidance requirement, communication reliability requirement, and environment dynamics relation yields the following stochastic optimization problem:

\vspace{-10pt}
\small
\begin{subequations}\label{eq:problem}
\begin{align}
\min_{\pi_s,\pi_p}
\quad&
\mathbb{E}_{\pi_s,\pi_p}
\left[
\sum_{k=1}^{H}
\ell_k
\right]
\label{eq:original_problem}
\\
\mathrm{s.t.}
\quad&
\Pr_{\pi_s,\pi_p}
\Big(
\mathbf{z}_{c}^{H}
\in
\mathcal{S}_{\mathrm{safe}}
(
\mathbf{z}_{u}^{H}
)
\Big)
\ge
1-\delta_1,
\label{Eq:safety_constraint0}
\\
&
\Pr_{\pi_s,\pi_p}
\Big(
\boldsymbol{\gamma}^{H}
\in
\mathcal{S}_{\mathrm{comm}}
\Big)
\ge
1-\delta_2,
\label{Eq:communications_constraint0}
\\
&
s_{k+1}\sim p_s(\cdot|s_k,a_{s,k}),
\qquad
k=1,\ldots,H,
\label{Eq:dynamics_constraint0}
\\
&
a_{s,k}\in\mathcal{A}_s,\quad
a_{p,k}\in\mathcal{A}_p,
\qquad
k=1,\ldots,H.
\label{Eq:action_constraints}
\end{align}
\end{subequations}
\normalsize

\noindent
Problem~\eqref{eq:problem} remains challenging due to the unknown stochastic environment dynamics, nonconvex chance constraints, and coupling between control and communication decisions.

\section{Policy Optimization in Latent Space}
This section introduces a tractable latent-space reformulation of Problem~\eqref{eq:problem} together with the corresponding policy optimization framework.

\subsection{Problem Reformulation}
Problem \eqref{eq:problem} is difficult to solve directly due to the trajectory-level chance constraints and the coupling between communication and control decisions. To facilitate policy optimization, the chance constraints are first reformulated and incorporated into the objective using barrier functions.

The safety and communication requirements are expressed as probabilities over entire trajectories. Define
\begin{subequations}
\begin{align}
P_{\mathrm{safe}}
&=
\Pr_{\pi_s,\pi_p}
\Big(
\mathbf z_c^H
\in
\mathcal{S}_{\mathrm{safe}}
(
\mathbf z_u^H
)
\Big),
\\
P_{\mathrm{comm}}
&=
\Pr_{\pi_s,\pi_p}
\Big(
\boldsymbol\gamma^H
\in
\mathcal{S}_{\mathrm{comm}}
\Big),
\end{align}
\end{subequations}
which denote the probabilities of satisfying the safety and communication requirements over the planning horizon, respectively. The expectations are taken with respect to the action policies.
These probabilities can be equivalently expressed using indicator functions as
\begin{subequations}
\begin{align}
P_{\mathrm{safe}}
&=
\mathbb E
\!\left[
\mathbb I_{\mathrm{safe}}
(
\mathbf z_c^H,
\mathbf z_u^H
)
\right],
\\
P_{\mathrm{comm}}
&=
\mathbb E
\!\left[
\mathbb I_{\mathrm{comm}}
(
\boldsymbol\gamma^H
)
\right],
\end{align}
\end{subequations}
where
\small
$\mathbb I_{\mathrm{safe}}
\!=\!
\mathbb I\left\{
\mathbf z_c^H
\!\in\!
\mathcal S_{\mathrm{safe}}(\mathbf z_u^H)
\right\}$
\normalsize
and
\small
$\mathbb I_{\mathrm{comm}}
\!=\!
\mathbb I\left\{
\boldsymbol\gamma^H
\!\in\!
\mathcal S_{\mathrm{comm}}
\right\}$\normalsize,

\noindent
with $\mathbb I\{\cdot\}$ denoting the indicator of an event. Accordingly, the chance constraints in \eqref{Eq:safety_constraint0} and \eqref{Eq:communications_constraint0} are equivalently written as $P_{\mathrm{safe}}\ge 1-\delta_1$ and $P_{\mathrm{comm}}\ge 1-\delta_2$, respectively.
The chance constraints are incorporated into the objective using logarithmic barrier functions, yielding

\vspace{-10pt}
\small
\begin{align}
\min_{\pi_s,\pi_p}
~
\mathbb E
\left[
\sum_{k=1}^{H}
\ell_k
\right]
\!-\!\frac{1}{\eta}
\log
\frac{\left(
P_{\mathrm{safe}}-1\!+\!\delta_1
\right)}{\left(
P_{\mathrm{comm}}-1\!+\!\delta_2
\right)^{-1}}
\label{eq:barrier_problem}
\end{align}
\normalsize

\noindent
where $\eta>0$ is the barrier parameter. Minimizing \eqref{eq:barrier_problem} seeks policies that achieve a low expected cost while satisfying the prescribed safety and communication-reliability requirements.

\subsection{Latent Rollout Generation}
The optimization problem in \eqref{eq:barrier_problem} requires estimating the expected cost, as well as the safety and communication reliability probabilities. Since these quantities depend on future latent trajectories, we employ Monte Carlo rollouts generated by the learned JEPA model.

Let $\theta_s$ and $\theta_p$ denote the parameters of the control and communication policies, respectively. For rollout $i=1,\ldots,K$ and planning step $k=1,\ldots,H$, the control and communication actions are sampled according to
\begin{subequations}
\begin{align}
a_{s,k}^{(i)}
&\sim
\pi_s^{\theta_s}
(
\cdot|
\bar z_k^{(i)}
),
\\
a_{p,k}^{(i)}
&\sim
\pi_p^{\theta_p}
(
\cdot|
\bar z_k^{(i)}
).
\end{align}
\end{subequations}

The unknown environment dynamics are approximated using the learned JEPA latent predictor introduced in \eqref{eq:jepa_predictor}. Accordingly, the latent trajectory is propagated according to
\begin{align}
z_{k+1}^{(i)}
\approx
f_\psi
\!\left(
\bar z_k^{(i)},
a_{s,k}^{(i)}
\right),
\label{eq:latent_rollout}
\end{align}
where the approximation reflects the prediction error of the learned latent dynamics model. By recursively feeding predicted latent states back into the predictor, future latent trajectories are generated over the planning horizon.

Repeating this procedure for $K$ rollouts produces a collection of latent trajectories that approximate future environment evolution under the current policies. These rollouts are subsequently used to estimate the expected cost together with the safety and communication reliability probabilities. As will be demonstrated in Section~IV, the resulting rollout prediction errors remain sufficiently small over the considered planning horizons.

\subsection{Differentiable Constraint Approximation}

Direct evaluation of the safety and communication chance constraints is generally non-differentiable. Therefore, we introduce differentiable surrogate models that enable gradient-based optimization. For safety evaluation, a neural-network-based collision-risk predictor
$
\hat p_{\mathrm{coll}}
(z_c,z_u)
\in [0,1]
$
is trained to estimate the probability of collision between controllable and uncontrollable entities represented in latent space. Specifically,
\begin{align}
\hat p_{\mathrm{coll}}
(z_c,z_u)
\approx
\Pr
\big(
d(\Omega_c,\Omega_u)
\le d_{\min}
\big),
\end{align}
where $\Omega_c$ and $\Omega_u$ denote the controllable and uncontrollable entities, respectively, $d(\Omega_c,\Omega_u)$ denotes the minimum boundary-to-boundary occupancy distance, and $d_{\min}$ is the collision threshold.

For rollout $i$, the trajectory-level safety score is approximated by

\vspace{-15pt}
\small 
\begin{align}
s_{\mathrm{safe}}^{(i)}
=
\mathrm{m}^{}_{\beta}
\left(\left\{1-
\hat p_{\mathrm{coll}}
(
z_{c,k}^{(i)},
z_{u,k}^{(i)}
) \right\}_{k=1}^{H}
\right),
\end{align}
\normalsize

\noindent
where the soft minimum operator $\mathrm{m}^{}_{\beta}(\cdot)$ with smoothing parameter $\beta>0$ is defined as

\vspace{-10pt}
\small
\begin{align}
\mathrm{m}^{}_{\beta}
(\{x_k\}_{k=1}^{H})
=
-\frac{1}{\beta}
\log
\left(
\sum_{k=1}^{H}
e^{-\beta x_k}
\right).
\end{align}
\normalsize

\noindent
The parameter $\beta$ controls the approximation to the minimum safety level over the planning horizon. A larger value of $\beta$ places greater emphasis on the time step with the highest predicted collision risk. The resulting rollout-based safety surrogate is
\vspace{-10pt}
\begin{align}
\hat P_{\mathrm{safe}}
=
\frac{1}{K}
\sum_{i=1}^{K}
s_{\mathrm{safe}}^{(i)}.
\end{align}

To evaluate communication reliability, we introduce a differentiable neural-network-based channel predictor
$
\hat h
(
z_c,
z_u,
s_{\mathrm{BS}}
)
\approx
h
(
\Omega_c,
\Omega_u,
s_{\mathrm{BS}}
),
$
which estimates the wireless channel coefficient from the latent representation and BS state. Using the predicted channel coefficient, the SNR-threshold violation indicator at time step $k$ for rollout $i$ is smoothly approximated by

\vspace{-10pt}
\small
\begin{align}
\hat p_{\mathrm{out},k}^{(i)}
=
\sigma_\alpha
\left(
1-
\frac{
|\hat h_k^{(i)}|^2a_{p,k}^{(i)}
}{
P_n\gamma_{\min}
}
\right),
\end{align}
\normalsize
where
\vspace{-10pt}
\begin{align}
\sigma_\alpha(x)
=
(1+e^{-\alpha x})^{-1}
\end{align}
denotes a sigmoid function with slope parameter $\alpha>0$.

To obtain a tractable differentiable surrogate, we replace the all-step communication event with a time-averaged relaxation. Accordingly, for rollout $i$, the communication-reliability score is

\vspace{-20pt}
\small
\begin{align}
s_{\mathrm{comm}}^{(i)}
=
\frac{1}{H}
\sum_{k=1}^{H}
\left(
1-
\hat p_{\mathrm{out},k}^{(i)}
\right),
\end{align}
\normalsize

\noindent
and its rollout-based estimate is

\vspace{-10pt}
\small
\begin{align}
\hat P_{\mathrm{comm}}
=
\frac{1}{K}
\sum_{i=1}^{K}
s_{\mathrm{comm}}^{(i)}.
\end{align}
\normalsize

\noindent
This relaxation penalizes the fraction of the horizon over which the SNR requirement is violated while permitting short violations. However, the
safety surrogate emphasizes the least-safe time step because a single collision can invalidate the trajectory. The resulting quantities
$\hat P_{\mathrm{safe}}$ and $\hat P_{\mathrm{comm}}$ are differentiable surrogates for trajectory safety and communication reliability and are
incorporated into the optimization objective.

\subsection{Sequential Policy Optimization}

Using the rollout-based estimates of the expected cost, safety probability, and communication reliability probability, the barrier-augmented optimization objective is given by

\vspace{-10pt}
\small
\begin{align}
J_\eta
(
\theta_s,
\theta_p
)
\!=\!
\frac{1}{K}
\sum_{i=1}^{K}
\sum_{k=1}^{H}
\ell_k^{(i)}
\!-\!\left(\varepsilon_b\!+\!\frac{1}{\eta}\right)
\log
\frac{\!\left(
\hat P_{\mathrm{safe}}
\!-\!1\!+\!\delta_1
\right)\qquad}{\!\left(
\hat P_{\mathrm{comm}}
\!-1\!+\!\delta_2
\right)^{-1}}
\label{eq:final_objective},
\end{align}
\normalsize

\noindent
where $\eta>0$ is the barrier parameter and $\varepsilon_b>0$ 
is a residual barrier weight that prevents the influence of the constraint terms from vanishing as $\eta$ increases.

The objective in \eqref{eq:final_objective} is differentiable with respect to the policy parameters $\theta_s$ and $\theta_p$. Consequently, gradients can be computed directly through automatic differentiation without requiring analytical gradient derivations. In principle, \eqref{eq:final_objective} can be optimized jointly with respect to the control and communication policies. However, direct joint optimization often suffers from unstable convergence due to the strong coupling between trajectory planning and communication power allocation. Furthermore, communication power can typically be adapted at a faster timescale than UAV motion, whereas vehicle dynamics impose stronger constraints on trajectory evolution.

Motivated by these observations, the proposed CRPL adopts a sequential optimization strategy. First, the control policy is optimized assuming the maximum allowable communication power, thereby prioritizing safe and goal-directed trajectory generation. Then, with the control policy fixed, the communication policy is optimized along the latent trajectories generated by the learned control policy. The resulting procedure consists of the following two steps:
\begin{itemize}
\item \textbf{Step 1:} Optimize the control policy $\pi_s^{\theta_s}$ assuming maximum communication power.
\item \textbf{Step 2:} Optimize the communication policy $\pi_p^{\theta_p}$ along the trajectories generated by the optimized control policy.
\end{itemize}
The proposed CRPL algorithm is summarized in Algorithm~1.

\begin{algorithm}[t]
\caption{CRPL Policy Optimization}
\label{alg:risk_tolerant}
\begin{algorithmic}[1]

\State \textbf{Input:}
initial policy parameters $\theta_s,\theta_p$,
barrier parameter $\eta>0$,
risk tolerances $\delta_1,\delta_2$,
planning horizon $H$,
number of rollouts $K$,
barrier growth factor $\mu>1$,
learning rate $\lambda>0$,
termination tolerances
$\epsilon_0,\epsilon_1,\epsilon_2>0$,
stabilization constant $\varepsilon_b > 0$.

\Repeat

\Statex \textbf{Stage 1: Control Policy Optimization}

\Repeat

\State Set:
$
a_{p,k}^{(i)} = a_{p,\max}
$

\State Generate $K$ latent rollouts:

\quad
$
a_{s,k}^{(i)}
\sim
\pi_s^{\theta_s}
(\cdot|\bar z_k^{(i)}),
$ \qquad
    $
    z_{k+1}^{(i)}
    \approx
    f_\psi
    (
    \bar z_k^{(i)},
    a_{s,k}^{(i)}
    )
    $

\State Compute
$\hat P_{\mathrm{safe}}$,
$\hat P_{\mathrm{comm}}$,
and
$J_\eta(\theta_s,\theta_p)$

\State Update control policy:
\[
\theta_s
\leftarrow
\theta_s
-
\lambda
\nabla_{\theta_s}
J_\eta
\]

\Until{
$
\|
\nabla_{\theta_s}
J_\eta
\|
<
\epsilon_1
$
}

\Statex \textbf{Stage 2: Communication Policy Optimization}

\Repeat

\State Generate $K$ latent rollouts:

\quad
$
a_{s,k}^{(i)}
\sim
\pi_s^{\theta_s}
(\cdot|\bar z_k^{(i)}),
$ \qquad
    $
    z_{k+1}^{(i)}
    \approx
    f_\psi
    (
    \bar z_k^{(i)},
    a_{s,k}^{(i)}
    )
    $
    
    \State
    $
    a_{p,k}^{(i)}
    \sim
    \pi_p^{\theta_p}
    (\cdot|\bar z_k^{(i)})
    $

\State Compute
$\hat P_{\mathrm{safe}}$,
$\hat P_{\mathrm{comm}}$,
and
$J_\eta(\theta_s,\theta_p)$.

\State Update communication policy:
\[
\theta_p
\leftarrow
\theta_p
-
\lambda
\nabla_{\theta_p}
J_\eta
\]

\Until{
$
\left\|
\nabla_{\theta_p}
J_\eta
\right\|
<
\epsilon_2
$
}

\State Update barrier parameter:
\[
\eta \leftarrow \mu\eta
\]

\Until{
$
\frac{1}{\eta}
<
\epsilon_0
$
}

\State \textbf{Output:}
optimized policy parameters
$\theta_s^\star$
and
$\theta_p^\star$.
\end{algorithmic}

\end{algorithm}

\section{Simulation Results}
\label{sec:Simulations}

In this section, we evaluate the proposed CRPL framework through numerical simulations in a synthetic UAV environment. We investigate the accuracy of the learned latent model, the resulting trajectory optimization performance, and the robustness of the framework under varying communication bandwidth and environment uncertainty levels.

\subsection{Simulation Setup}
\label{sec:sim_setup}

\subsubsection{Environment and Dynamics}
\label{sec:env_dynamics}

We consider a synthetic two-dimensional square environment $[0,1]\times[0,1]$ in normalized spatial units. The UAV starts from $r_{c,0}=[0.1,0.5]^T$ with zero initial velocity and aims to reach a circular goal region centered at $r_g=[0.9,0.5]^T$ with radius $0.05$. A BS located at $r_{\mathrm{BS}}=[0.5,0.05]^T$ communicates with and controls the UAV through a wireless link. A planning horizon of $H=100$ time steps is adopted based on the JEPA's look-ahead evaluation presented later.

Let
$
s_{c,k}
\!=\!
\begin{bmatrix}
r_{c,k}^{T},
v_{c,k}^{T}
\end{bmatrix}^{T}
$
denote the UAV state, where $r_{c,k},v_{c,k}\in\mathbb R^2$ are its position and velocity, respectively. The second-order dynamics can be written in the state-space form
\begin{align}
{s}_{c,k+1}
=
A s_{c,k}
+
B \tilde{a}_{s,k}
+
G w_k,
\label{eq:simulation_second_order}
\end{align}
where
$
A
\!=\!
\begin{bmatrix}
I & I\\
0 & I
\end{bmatrix},
B
\!=\!
\begin{bmatrix}
0\\
I
\end{bmatrix},
G
\!=\! 
\begin{bmatrix}
0\\
{\sigma}_{c} I
\end{bmatrix}
$,
$I$ denotes the two-dimensional identity matrix, 
$
w_k
\!\sim\!
\mathcal N
\left(
0,
I
\right)
$
models the disturbance, $\sigma_c$ controls the level of dynamic uncertainty, the speed $v_{c,k}$ is limited to $v_{\max} \! = \!4\!\times\!10^{-2}$ before the UAV position is updated,
and
$
\tilde{a}_{s,k}
=
a_{\max}a_{s,k},
$
is the physical acceleration command, where $a_{\max} \! = \! 3.5\!\times\!10^{-3}$ is the maximum acceleration magnitude.
For closed-loop control, the normalized motion-action space is
$
\mathcal A_s
\!=\!
\left\{
[0, 0],
[1, 0],
[-1, 0],
[0, 1],
[0, -1]
\right\},
$
and the communication-power action space is
$
\mathcal A_p
\!=\!
\{1,4\}.
$
During policy optimization, differentiable continuous relaxations of the motion and communication actions are used to propagate gradients through the
latent rollouts. At execution time, the relaxed policy outputs are mapped to the nearest feasible actions in $\mathcal A_s$ and $\mathcal A_p$,
respectively. Accordingly, the UAV may apply zero acceleration or accelerate along one of the positive or negative coordinate directions, while the communication policy selects between the two available transmission power levels. During offline JEPA data generation, the two normalized acceleration components are instead sampled continuously from $[-1,1]$ to provide broader excitation of the second-order dynamics. The closed-loop trajectory experiments include one obstacle fixed at $r_{u,0}$, which depends on the scenario under consideration. However, for latent model training, the uncontrollable entity evolves according to
$
r_{u,k+1}
\!=\!
r_{u,k}
+
v_{u,k},
$
where its direction is sampled independently for every trajectory with deterministic reflection at the environment boundaries, and with a fixed speed set as $\|v_{u,k}\|_2 \!=\! 1.25 \times 10^{-2}$.

\subsubsection{JEPA Architecture and Training}
\label{sec:jepa_architecture}

The JEPA encoder $\phi(\cdot)$ maps each $100\!\times\!100$ observation image $I_k$ to a compact latent representation. It consists of four convolutional
blocks with channel dimensions $(32,64,96,128)$, followed by adaptive pooling and a fully connected projection to a $256$-dimensional feature representation. Two latent heads then generate the controllable and uncontrollable components $z_{c,k},z_{u,k}\!\in\!\mathbb R^{16}$, whose concatenation forms the complete latent state $z_k\!\in\!\mathbb R^{32}$.

Although the general formulation supports an arbitrary history length $M$, we use $M\!=\!2$ in the simulations. For each entity $j\!\in\!\{c,u\}$, define the augmented latent state
$
\bar z_{j,k}
\!=\!
\begin{bmatrix}
z_{j,k}^{T},
z_{j,k-1}^{T}
\end{bmatrix}^{T}
$.
The corresponding linear latent predictor is parameterized as
\begin{align}
\bar z_{j,k+1}
\!\approx\!
\begin{bmatrix}
\psi_{j,0} & \psi_{j,1}\\
I & 0
\end{bmatrix}
\bar z_{j,k}
+
\begin{bmatrix}
\psi_{j,a}\\
0
\end{bmatrix}
a_{j,k},
\label{eq:entity_latent_predictor}
\end{align}
where
$
\psi_{j,0},\psi_{j,1}\in\mathbb R^{16\times16}
$
and
$
\psi_{j,a}\in\mathbb R^{16\times2}
$
are learned parameters. The controllable predictor receives $a_{c,k}\!=\!a_{s,k}$, whereas the uncontrollable predictor receives no
exogenous action and therefore uses $a_{u,k}\!=\!0$. The inclusion of both $z_{j,k}$ and $z_{j,k-1}$ enables the predictor to capture temporal
dependencies while maintaining a lightweight model structure. The controllable and uncontrollable predictors use separate parameters, allowing
their dynamics to be learned independently.

The JEPA model is trained offline using an exponential-moving-average target encoder with momentum $\tau\!=\!0.995$. After training, the encoder and latent predictors are fixed and used to generate recursive latent rollouts during CRPL policy optimization. To isolate the impact of predictive latent planning, the linear predictor in \eqref{eq:entity_latent_predictor} is employed in the main closed-loop experiments. More expressive predictors can be incorporated without modifying the proposed optimization framework; accordingly, a deep multilayer perceptron (MLP) and a long short-term memory (LSTM) network are evaluated separately as alternative predictor architectures in the latent rollout comparison.

\subsubsection{Wireless Channel}
\label{sec:comm_channel}
Communication performance is evaluated using a geometry-aware LOS/NLOS channel model. The large-scale LOS channel gain is modeled according to the distance dependent path loss model
$
|h_k^{\mathrm{LOS}}|^2\!=\!G_0 d_k^{-\alpha_\mathrm{PL}},
$
where $G_0\!=\!1$ is the reference gain, $\alpha_\mathrm{PL}\!=\!2$ is the path-loss exponent, and
$d_k\!=\!\|r_{c,k}-r_{\mathrm{BS}}\|_2$
denotes the distance between the UAV and the BS. To account for obstacle-induced shadowing, we employ a simplified LOS/NLOS approximation. The effective channel gain is given by
$
|h_k|^2\!=\!(1-p_k)|h_k^{\mathrm{LOS}}|^2+p_k\kappa_{\mathrm{NLOS}}|h_k^{\mathrm{LOS}}|^2,
$
where $p_k\!\in\![0,1]$ denotes the blockage probability, and $\kappa_{\mathrm{NLOS}}\!=\!0.1$ represents the additional attenuation associated with NLOS propagation. The blockage probability $p_k$ is computed from the relative geometry of the UAV, obstacle, and BS using a smooth differentiable approximation of line-of-sight obstruction. As a result, the channel quality degrades smoothly as the obstacle increasingly obstructs the LOS path between the UAV and the BS.

\subsubsection{Communication Protocol}
\label{sec:comm_protocol}
The wireless channel model is used to evaluate closed-loop communication-control performance through a packet-delivery model. The CRPL policies are optimized using the fixed SNR requirement $\gamma_{\min}$ and are not explicitly conditioned on $B$; bandwidth variation is introduced only during closed-loop packet-delivery evaluation. Since the achievable transmission rate is determined by the received SNR, packet delivery provides an operational measure of the communication reliability formulation introduced in Section~\ref{sec:system_model}. At each time step, the BS computes a control command and transmits a control packet of size $L_{\mathrm{pkt}} = 4$ bits to the UAV. The achievable communication rate is given by
$
R_k
=
B
\log_2(1+\gamma_k),
$
where $B$ denotes the available bandwidth and $\gamma_k$ is the received SNR. A communication outage is declared whenever the available channel capacity is insufficient to deliver the control packet within the current control interval, i.e.,
$
R_k\Delta t < L_{\mathrm{pkt}}.
$
During an outage, the UAV continues to apply the most recently received control action until a new control packet is successfully delivered. The outage duration $T_{\mathrm{out}}$ reported in Figs.~\ref{fig:trade_offs_bandwidth} and \ref{fig:trade_offs_sigmas} is computed as the cumulative time for which this condition holds. This protocol is employed only in the experiments of Section~\ref{sec:trade_offs} to evaluate the communication-control system under bandwidth limitations and environment uncertainty.

\subsubsection{Policy and Algorithm Parameters}
\label{sec:policy_algo_params}
The control and communication policies, $\pi_s^{\theta_s}$
and
$\pi_p^{\theta_p}$, are parameterized by feed-forward neural networks with two hidden layers of $64$ neurons and $\tanh$ activations. Both networks take the latent representation as input and output action probabilities over $\mathcal A_s$ and $\mathcal A_p$, respectively.
Unless otherwise stated, the default simulation parameters are: collision threshold $d_{\min}\!=\!0$, safety risk tolerance $\delta_1\!=\!0.05$,
communication outage tolerance $\delta_2=0.05$, minimum required SNR $\gamma_{\min}\!=\!1$~($0$~dB), barrier growth factor $\mu\!=\!2$, initial barrier parameter $\eta\!=\!1$, barrier stabilization parameter $\varepsilon_b\!=\!10^{-4}$, learning rate $\lambda\!=\!10^{-3}$, barrier termination tolerance $\epsilon_0\!=\!10^{-3}$, control-policy convergence tolerance $\epsilon_1\!=\!10^{-3}$, and communication-policy convergence tolerance $\epsilon_2\!=\!10^{-3}$.

\subsubsection{Evaluation Metrics}
\label{sec:metrics}

The terminal goal-reaching error is defined as
$
d_{\mathrm{f}}
=
\left\|r_{c,H}-r_g\right\|_2,
$
where $r_{c,H}$ denotes the terminal UAV position. The cumulative communication energy is approximated by
$
E_{\mathrm{comm}}
=
\sum_{k=1}^{H} a_{p,k},
$
which represents the total normalized transmission-power expenditure over the planning horizon.
The UAV propulsion energy is evaluated as
$
E_{\mathrm{ctrl}}
=
\sum_{k=1}^{H}P\!\left(\left\|v_{c,k}\right\|_2\right),
$
where $P(\cdot)$ follows the standard rotary-wing propulsion-power model in \cite{Zeng2019rotary}, which accounts for profile, induced, and parasite
power components. The adopted propulsion parameters are $P_0=79$, $P_i=88$, $U_{\mathrm{tip}}=120$, $v_0=4$, $d_0=0.3$, $\rho=1.225$, $s=0.05$, and $A=0.5$. Safety performance is quantified by the maximum collision probability
$
p_{\mathrm{coll}}^{\max}
=
\max_{k=1,\ldots,H}p_{\mathrm{coll},k},
$
where $p_{\mathrm{coll},k}$ denotes the instantaneous ground-truth collision probability. Communication reliability is measured by the outage duration
$T_{\mathrm{out}}$, defined as the cumulative time during which the outage condition in Section~\ref{sec:comm_protocol} is satisfied.

\subsubsection{Proposed and Baseline Methods}
\label{sec:baselines}

We compare the \textbf{CRPL} framework against three baseline methods. \textbf{APC} is an analytical predictive controller with access to the
environment dynamics and obstacle evolution, and therefore serves as an oracle benchmark that is not realizable in practice. \textbf{RCC} is a
reactive constrained controller that employs the same collision-avoidance constraints as CRPL together with SNR-triggered transmission power adaptation
based only on current observations. \textbf{UCC} is an unconstrained communication controller that performs local collision avoidance and
reactive transmission power adaptation without incorporating communication or probabilistic safety constraints into the planning process.

\begin{figure}[t]
\centering

\includegraphics[
    width=0.95\linewidth
]{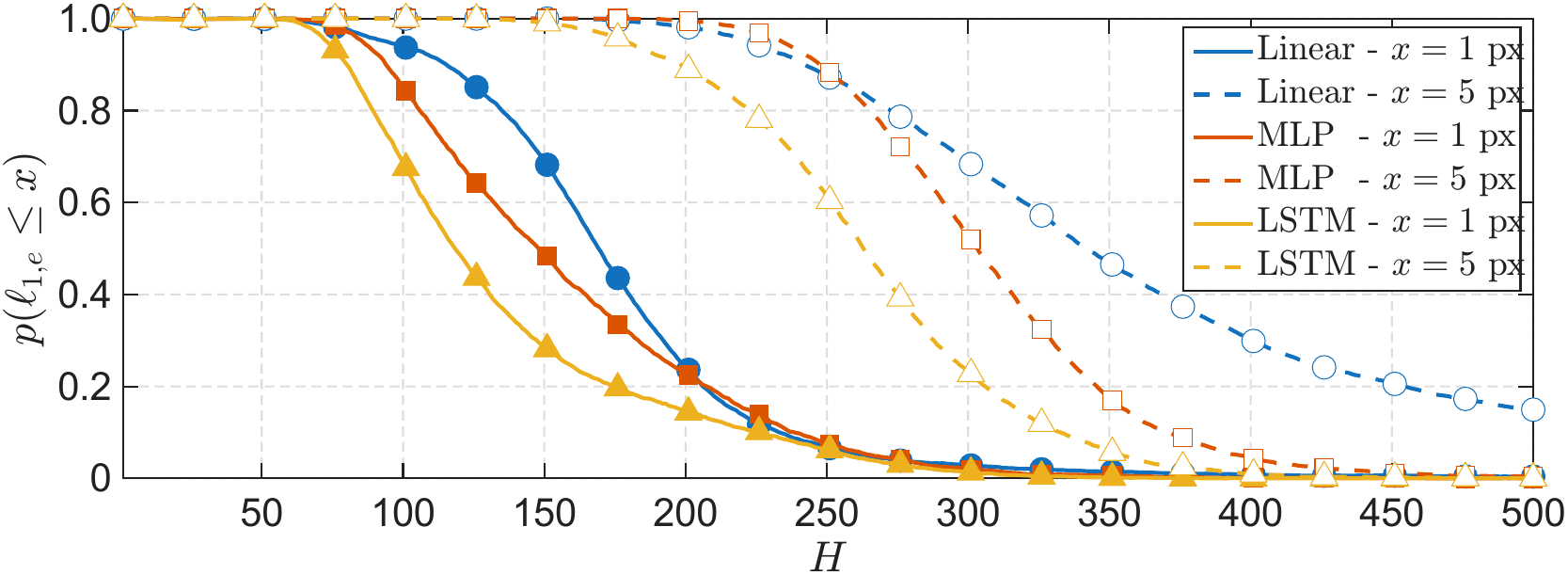}

\vspace{+0.2em}

\includegraphics[
    width=0.95\linewidth
]{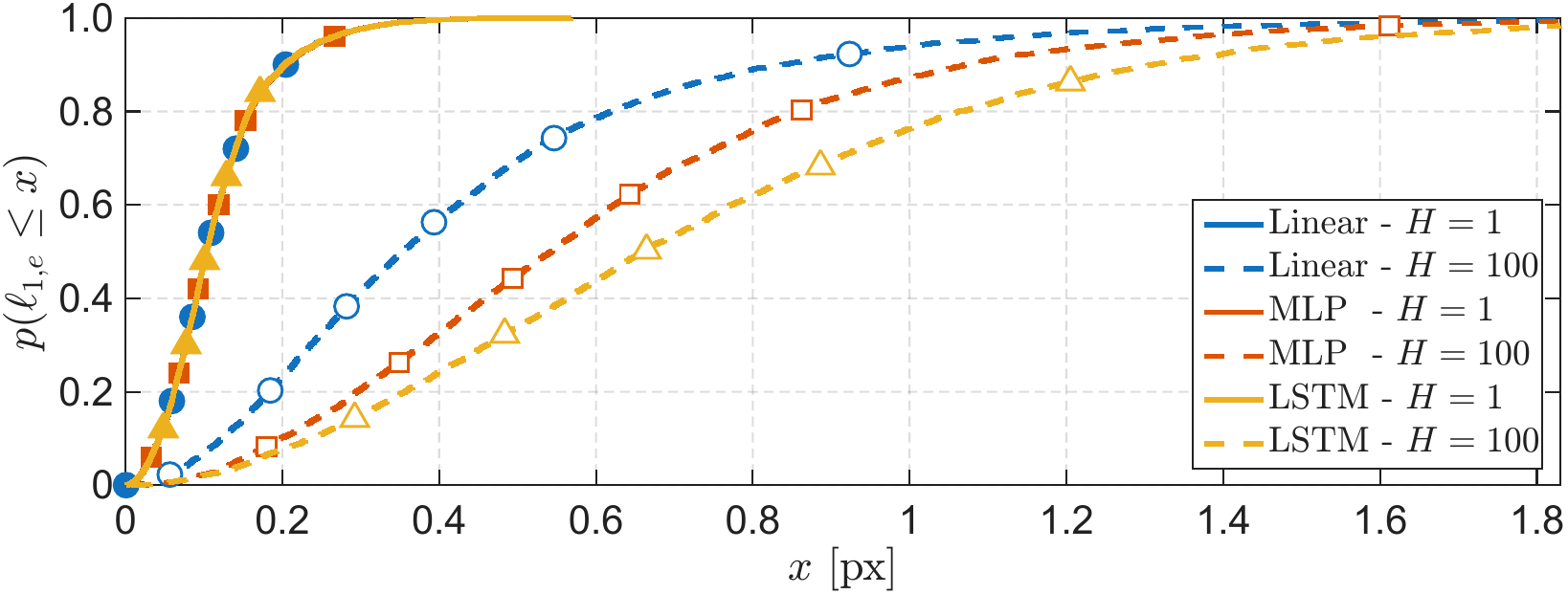}

\vspace{-0.2em}
\caption{Recursive controllable-latent predictor comparison.
(a) Probability that UAV position error remains below $x\in\{1,5\}$ pixels as a function of the look-ahead horizon $H$.
(b) Empirical CDF of the decoded $\ell_1$ position error at $H\!=\!1$ and
$H\!=\!100$.}
\label{fig:rollout_error_combined}

\vspace{-1.0em}
\end{figure}

\subsection{Latent Rollout Accuracy}
\label{sec:world_model_performance}
Before evaluating the closed-loop performance of CRPL, we examine whether the learned JEPA representation supports stable recursive prediction and whether increasing the predictor complexity improves long-horizon accuracy. Accurate multi-step rollouts are essential because the proposed framework uses them to anticipate future UAV motion, communication quality, and collision risk over the planning horizon.

To isolate the effect of predictor architecture, the same frozen JEPA representation is used for all models, and only the controllable latent predictor is varied. We compare the proposed linear second-order predictor with a deep MLP and an LSTM network over a common set of Monte Carlo
trajectories. At the look-ahead horizon $H$, the decoded UAV position error is defined as
$
\ell_{1,e}(H)
=
\left\|
\hat r_{c,H}
-
r_{c,H}
\right\|_1,
\label{eq:rollout_position_error}
$
where $\hat r_{c,H}$ and $r_{c,H}$ denote the predicted and ground-truth UAV positions, respectively. The error is measured in image pixels, with one
pixel corresponding to $0.01$ normalized spatial units.

\figurename{~\ref{fig:rollout_error_combined}}(a) reports probability $p(\ell_{1,e}(H)\leq x)$ for thresholds $x\in\{1,5\}$ pixels. The
$1$-pixel threshold provides a stringent measure of local prediction accuracy, whereas the $5$-pixel threshold represents a task-relevant spatial
tolerance. Since five pixels correspond to $0.05$ normalized spatial units, this threshold is comparable to the radius of the goal region, although the rollout error and goal-reaching condition use different norms. All three predictors remain accurate over short horizons, but their performance
separates as recursive errors accumulate. At the planning horizon $H\!=\!100$, approximately $93\%$ of the linear-predictor rollouts remain below a
$1$-pixel error, compared with about $84\%$ for the MLP and $68\%$ for the LSTM. In contrast, more than $99\%$ of the rollouts for all three predictors
remain below the $5$-pixel threshold at $H\!=\!100$, indicating that large task-scale prediction errors are rare over the selected planning horizon.
The linear predictor nevertheless exhibits the slowest probability decay as the time horizon increases.

\begin{figure*}[t]
\centering
\begin{minipage}[t]{0.45\textwidth}
    \centering
    \includegraphics[
        width=0.9\linewidth,
        trim=0cm 0cm 0cm 0cm,
        clip
    ]{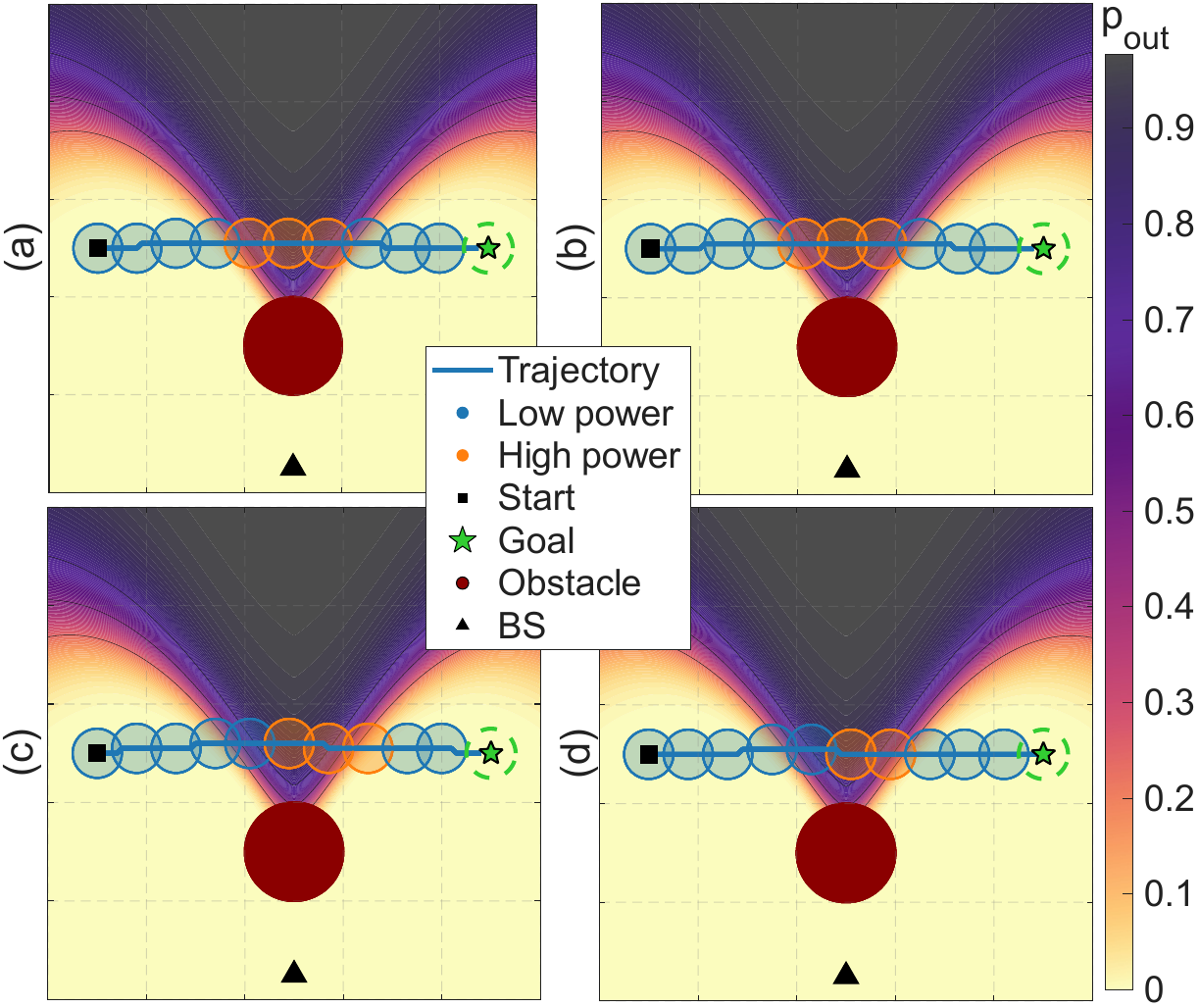}

    \vspace{0.3em}

    \small\textbf{Scenario 1}
\end{minipage}
\hfill
\begin{minipage}[t]{0.45\textwidth}
    \centering
    \includegraphics[
        width=0.9\linewidth,
        trim=0cm 0cm 0cm 0cm,
        clip
    ]{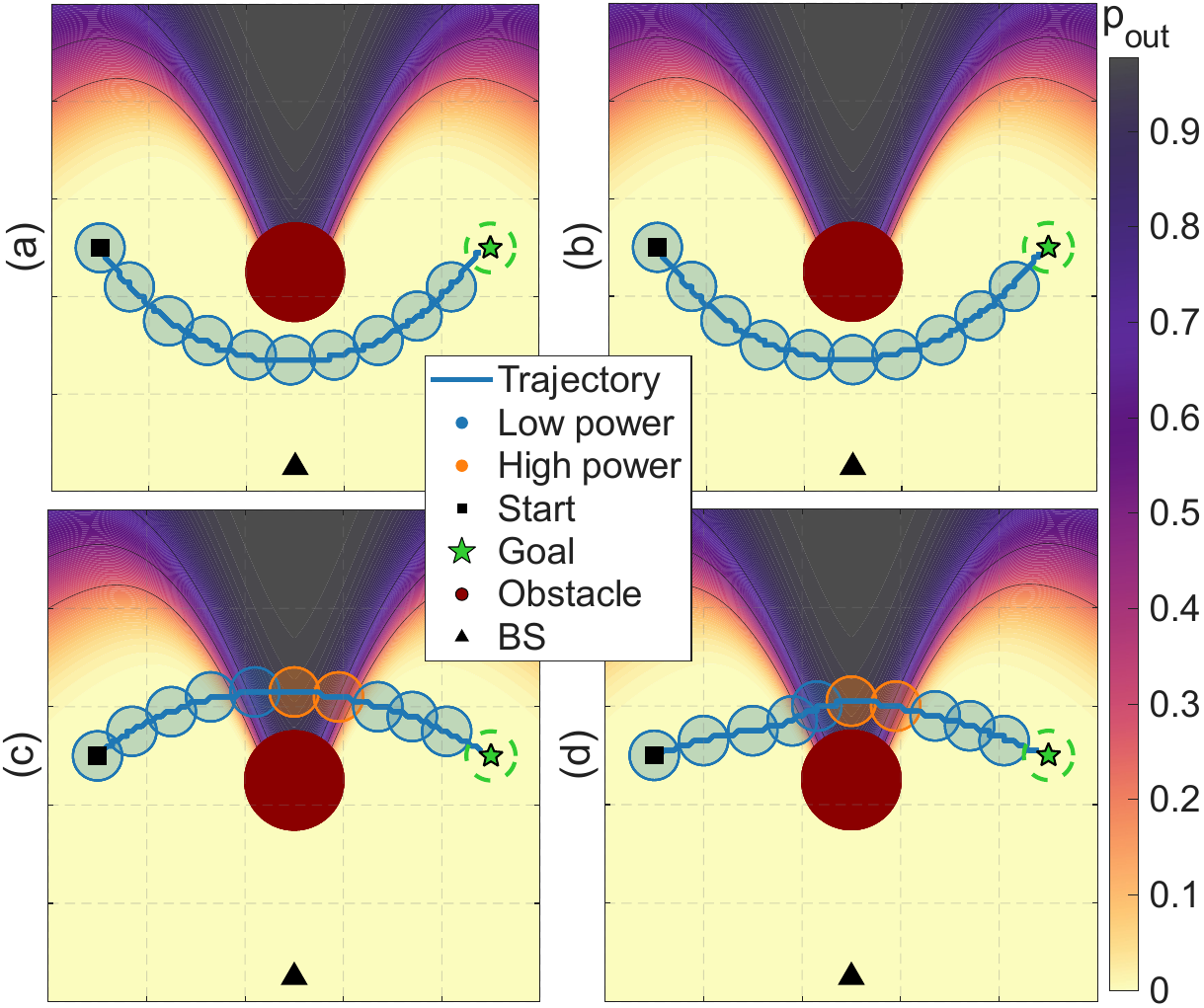}

    \vspace{0.3em}

    \small\textbf{Scenario 2}
\end{minipage}

\caption{UAV trajectories under two obstacle configurations. In Scenario~1, the obstacle is located at $(0.50,0.30)$, whereas in Scenario~2
it is located at $(0.50,0.45)$. Within each scenario, panels (a)--(d) correspond to CRPL, APC, RCC, and UCC, respectively. The background denotes
the geometry-dependent blockage probability, and the trajectory markers indicate the selected low- and high-power actions.}

\label{fig:Trajectories01}
\end{figure*}

\begin{figure*}[t]
    \centering
    
    \begin{minipage}[t]{0.47\textwidth}
        \centering
        
        \includegraphics[
            width=\linewidth,
            trim=0cm 0.0cm 0cm 0cm,
            clip
        ]{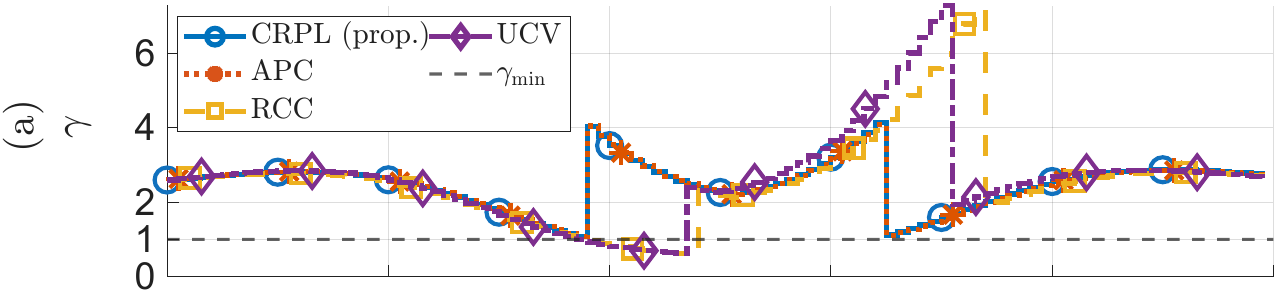}

        \vspace{0.4em}

    \end{minipage}
    \hfill
    \begin{minipage}[t]{0.47\textwidth}
        \centering
        
        \includegraphics[
            width=\linewidth,
            trim=0cm 0.0cm 0cm 0cm,
            clip
        ]{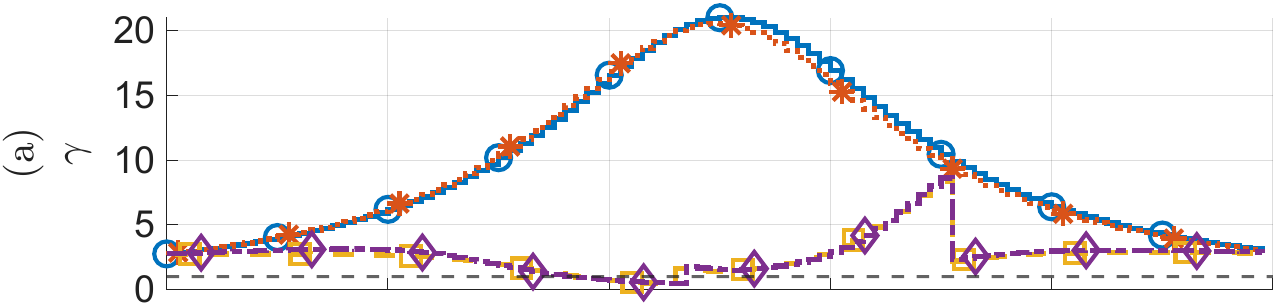}

        \vspace{0.4em}

    \end{minipage}

    \vspace{0.0em}

    \begin{minipage}[t]{0.47\textwidth}
        \centering
        
        \includegraphics[
            width=\linewidth,
            trim=0cm 0cm 0cm 0.0cm,
            clip
        ]{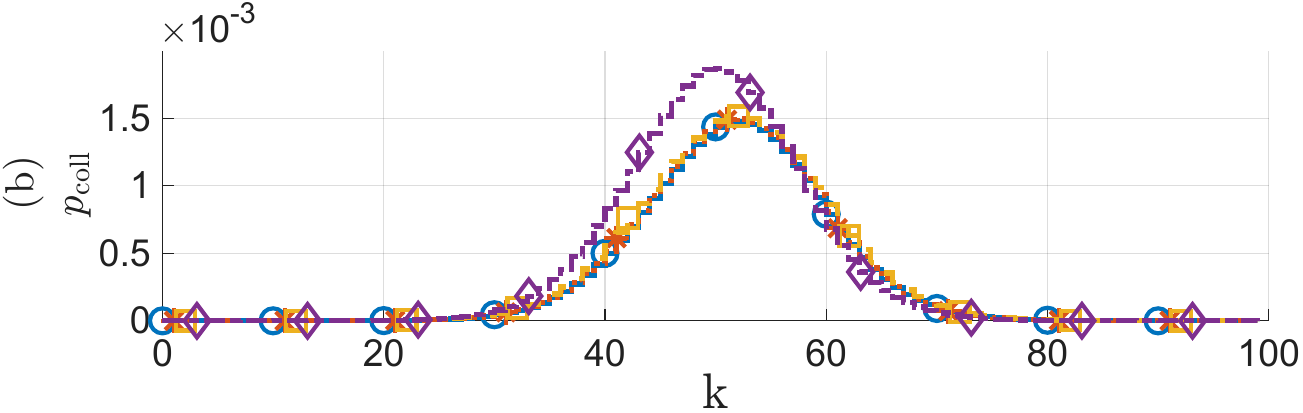}

        \vspace{0.4em}

        \small\textbf{Scenario 1}
    \end{minipage}
    \hfill
    \begin{minipage}[t]{0.47\textwidth}
        \centering
        
        \includegraphics[
            width=\linewidth
        ]{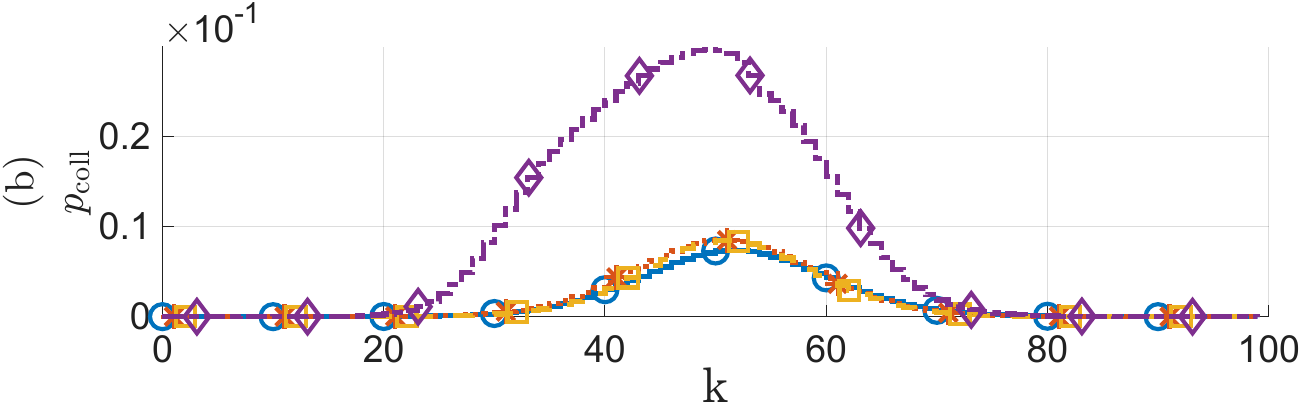}

        \vspace{0.4em}

        \small\textbf{Scenario 2}
    \end{minipage}

\caption{Received SNR and collision-risk evolution under the two obstacle configurations. Panel~(a) shows the received SNR $\gamma_k$, with the dashed
line indicating the required level $\gamma_{\min}$, while panel~(b) shows the corresponding collision probability $p_{\mathrm{coll},k}$.}

    \label{fig:snr_collision_combined}
\end{figure*}

\figurename{~\ref{fig:rollout_error_combined}}(b) presents the empirical error CDFs at $H\!=\!1$ and $H\!=\!100$. At one-step prediction, the three distributions are nearly coincident, with approximately $90\%$ of the errors remaining below $0.2$ pixels. At the planning horizon $H\!=\!100$, their tail behavior becomes more distinct. The $90$th-percentile errors are approximately $0.82$, $1.08$, and $1.28$ pixels for the linear, MLP, and LSTM predictors, respectively. Thus, to cover $90\%$ of the evaluated rollouts, the error thresholds required by the MLP and LSTM are approximately $1.3$ and $1.6$ times that of the linear predictor. This further demonstrates the greater stability of the linear model under recursive prediction.

These results indicate that, for the learned JEPA representation and the approximately second-order UAV dynamics, increasing predictor complexity
does not improve recursive accuracy. The lightweight linear predictor provides the most stable long-horizon rollouts and is therefore used in the
subsequent CRPL experiments. Its accuracy over the selected planning horizon supports the use of latent rollouts for evaluating future communication and
safety conditions without direct access to analytical environment dynamics.

\subsection{Trajectory Optimization Performance}
\label{sec:trajectory}

We evaluate the proposed CRPL framework under the two obstacle configurations shown in \figurename{~\ref{fig:Trajectories01}}, with the
corresponding SNR and collision-risk profiles presented in \figurename{~\ref{fig:snr_collision_combined}}. In Scenario~1, the obstacle is
located at $r_{u,0}=(0.50,0.30)$ and mainly degrades the communication channel along the direct path between the initial position and the goal. In
Scenario~2, the obstacle is located at $r_{u,0}=(0.50,0.45)$ and directly obstructs the shortest path. The two configurations therefore induce
different trade-offs among path efficiency, communication quality, and collision safety.

In Scenario~1, all methods remain close to the direct path and exhibit similar collision-risk profiles. The main difference lies in the timing of
the transmission-power decisions. CRPL and APC anticipate the upcoming channel degradation and increase power before the received SNR reaches
$\gamma_{\min}$. By contrast, RCC and UCC increase power only after sensing the degradation, causing a temporary SNR violation. Their power reduction is
also delayed, so high transmission power remains active after the strongest shadowing has passed, resulting in unnecessary communication-energy
consumption. The peak collision risk of UCC is only about $1.2$ times that of CRPL, reflecting the relatively mild geometric conflict in this scenario.

\begin{figure*}[t]
\centering
\includegraphics[
width=\linewidth,
trim=0cm 0cm 0cm 0cm,
clip
]{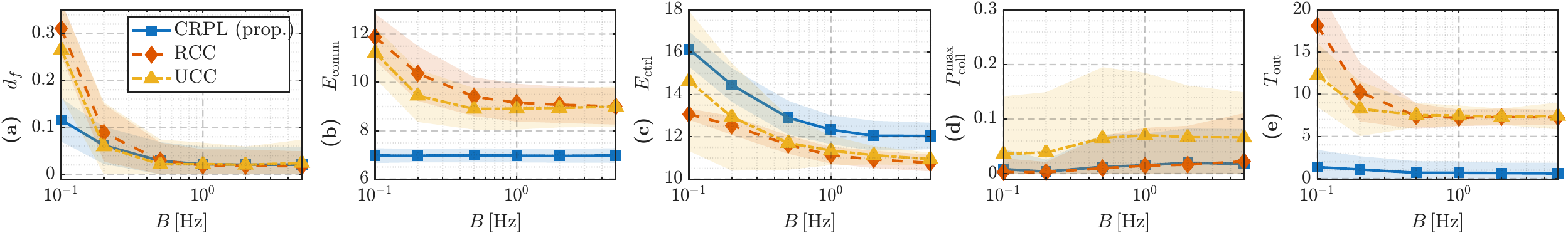}
\vspace{-2.0em}
\caption{Closed-loop performance versus communication bandwidth $B$.
Panels (a)--(e) show the terminal goal-reaching error $d_{\mathrm f}$,
communication energy $E_{\mathrm{comm}}$, motion energy
$E_{\mathrm{ctrl}}$, maximum collision probability
$p_{\mathrm{coll}}^{\max}$, and outage duration $T_{\mathrm{out}}$,
respectively.}
\label{fig:trade_offs_bandwidth}
\vspace{-0.5em}
\end{figure*}

\begin{figure*}[t]
\centering
\includegraphics[
width=\linewidth,
trim=0cm 0cm 0cm 0cm,
clip
]{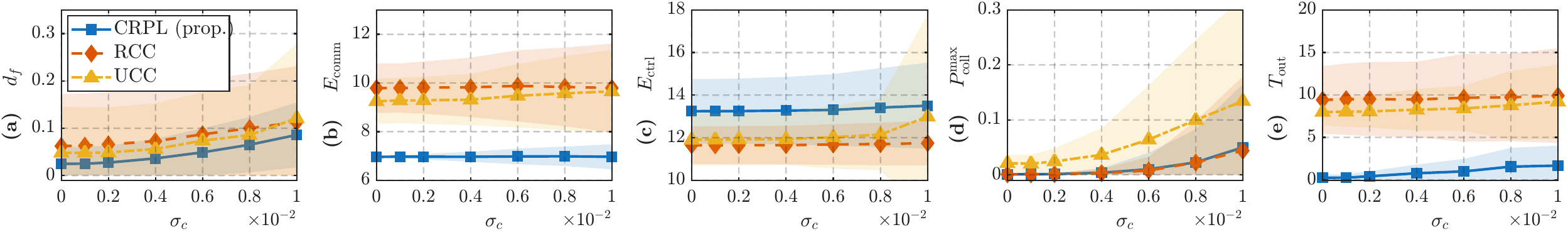}
\vspace{-2.0em}
\caption{Closed-loop performance versus dynamic disturbance $\sigma_c$.
Panels (a)--(e) show the terminal goal-reaching error $d_{\mathrm f}$,
communication energy $E_{\mathrm{comm}}$, motion energy
$E_{\mathrm{ctrl}}$, maximum collision probability
$p_{\mathrm{coll}}^{\max}$, and outage duration $T_{\mathrm{out}}$,
respectively.}
\label{fig:trade_offs_sigmas}
\end{figure*}

In Scenario~2, the benefit of predictive planning becomes more pronounced. CRPL and APC select a lower detour around the obstacle and thereby avoid the
more strongly shadowed region, while RCC and UCC follow the upper side of the obstacle and experience a deeper SNR degradation. Consequently, CRPL
and APC maintain the received SNR above the required level without sustained high-power transmission, whereas RCC and UCC temporarily fall below the
threshold before their delayed power increase takes effect. The collision-risk profiles show a similar separation: the peak risk of UCC is
approximately $4$ times that of CRPL, while APC and RCC remain much closer to the proposed method. This larger difference results from the absence of
a trajectory-level safety constraint in UCC.

The close agreement between CRPL and APC indicates that the learned latent representation preserves the information required for predictive,
communication-aware trajectory planning. Overall, CRPL anticipates future channel degradation and collision risk, allowing it to adapt both trajectory
and transmission power at the appropriate time and thereby achieve more reliable and efficient operation than the reactive alternatives.

\subsection{Limited Bandwidth and Dynamic Uncertainty}
\label{sec:trade_offs}

We next examine whether the observed
advantages of CRPL persist over a broader range of communication and dynamic conditions. In particular, we evaluate whether predictive latent planning remains effective when limited bandwidth constrains command delivery and dynamic uncertainty makes future UAV motion and channel evolution less predictable.

\figurename{~\ref{fig:trade_offs_bandwidth}} shows that the largest performance differences arise in the bandwidth-limited regime. At the lowest considered bandwidth, the terminal errors $d_{\mathrm f}$ of RCC and UCC are approximately $3$ and $2.5$ times that of CRPL, respectively, while their
communication energies $E_{\mathrm{comm}}$ are approximately $1.7$ and $1.6$ times that of CRPL. As the bandwidth increases, the terminal errors of all methods decrease and approach a saturation region, indicating that additional communication resources provide diminishing benefits once bandwidth is no longer the dominant performance bottleneck. The strongest separation is observed in the outage duration $T_{\mathrm{out}}$, for which RCC and UCC experience approximately $18$ and $12$ times the outage duration of CRPL, respectively, at the lowest bandwidth, and remain more than $10$ times that of CRPL over much of the considered range.

These trends are consistent with the trajectory, SNR, and collision-risk profiles observed in the preceding subsection. The recursive JEPA rollouts
allow CRPL to identify communication-shadowed and high-risk regions before they are reached and to coordinate trajectory deviations with transmission-power decisions. The reactive methods instead respond only after unfavorable channel or geometric conditions have been encountered, resulting in higher $E_{\mathrm{comm}}$ and prolonged outages. CRPL and RCC maintain low $p_{\mathrm{coll}}^{\max}$ because both enforce the safety constraint, whereas UCC exhibits a larger and more variable collision risk, consistent with the peaks observed in \figurename{~\ref{fig:snr_collision_combined}}. CRPL therefore combines the safety benefits of constrained control with the communication benefits of predictive planning, at the cost of only a moderate increase in $E_{\mathrm{ctrl}}$, approximately $1.1$--$1.2$ times that of the reactive controllers.

\figurename{~\ref{fig:trade_offs_sigmas}} shows the corresponding sensitivity to dynamic uncertainty. For small and moderate values of $\sigma_c$, the
performance changes gradually, indicating that all controllers tolerate limited deviations from the nominal UAV dynamics. The separation becomes
more pronounced at $\sigma_c=0.01$, where stronger disturbances reduce the accuracy of future motion and channel predictions. At this uncertainty
level, the terminal errors $d_{\mathrm f}$ of RCC and UCC are approximately $1.35$ and $1.4$ times that of CRPL, respectively, while their communication
energies $E_{\mathrm{comm}}$ are about $1.4$ times that of CRPL. Their outage durations $T_{\mathrm{out}}$ are approximately $7$ and $6.5$ times that of
CRPL, demonstrating that predictive trajectory and communication adaptation remains effective even when the realized UAV motion deviates from its
predicted evolution. The comparatively contained variation in the CRPL outage results also indicates more consistent communication reliability
under increasing uncertainty.

The comparison among CRPL, RCC, and UCC clarifies the contributions of the main algorithmic components. CRPL and RCC retain relatively low
$p_{\mathrm{coll}}^{\max}$ because both explicitly enforce the probabilistic safety constraint. In contrast, the collision risk of UCC increases more
rapidly with $\sigma_c$ and becomes approximately $3$ times that of CRPL at the highest disturbance level, while its larger variation indicates less
consistent safety performance under strongly perturbed motion. The substantially lower $d_{\mathrm f}$, $E_{\mathrm{comm}}$, and
$T_{\mathrm{out}}$ of CRPL relative to RCC isolate the additional benefit of predictive latent planning, since both methods retain the same safety
mechanism. Across both bandwidth and uncertainty studies, the comparatively contained variation of CRPL in terminal error and outage duration further
indicates more reliable closed-loop performance under unfavorable operating conditions.

Together with the rollout study in Section~\ref{sec:world_model_performance}, these results show that the stable look-ahead predictions provided by the JEPA representation and lightweight linear predictor translate into effective trajectory and communication adaptation. CRPL therefore achieves improved communication reliability, goal-reaching accuracy, and safety with only a moderate increase in motion energy $E_{\mathrm{ctrl}}$.

\section{Conclusion}
\label{sec:conclusion}

This article presented CRPL, a communication-aware and safety-aware predictive latent control framework for UAV trajectory optimization. By integrating a JEPA-based predictive model with probabilistic communication and safety constraints, CRPL jointly optimizes trajectory and communication decisions from high-dimensional observations. The proposed framework enables proactive adaptation to communication degradation and collision risk through predictive latent rollouts, thereby improving decision-making under uncertainty. Simulation results demonstrated consistent improvements in trajectory
accuracy, communication efficiency, reliability, and safety compared with reactive baseline methods, while limiting the additional motion effort required for communication-aware and collision-safe trajectory adaptation.

\balance

\ifCLASSOPTIONcaptionsoff
  \newpage
\fi


\end{document}